\documentclass[12pt]{article} \textheight=24 true cm
\usepackage{graphicx}
\usepackage{amsmath}
\usepackage{amssymb}
\begin{document}
\begin{center}
{\Large\bf Alfv\'{e}n wave in Higher Dimensional space time
}\\[8 mm]
D. Panigrahi\footnote{Relativity and Cosmology Research Centre,
Jadavpur University, Kolkata - 700032, India, e-mail:
dibyendupanigrahi@yahoo.co.in, Permanent Address : Kandi Raj
College, Kandi, Murshidabad 742137, India},
 Ajanta Das\footnote{Relativity and Cosmology Research Centre, Jadavpur
University, Kolkata - 700032, India,  Permanent Address : Heritage
Institute of Technology, Anandapur, Kolkata - 700107, India,
e-mail : ajanta.das@heritageit.com},
  and S. Chatterjee\footnote{Relativity and Cosmology Research Centre, Jadavpur University,
Kolkata - 700032, India, and also at NSOU, New Alipore College,
Kolkata  700053,
e-mail : chat\_ sujit1@yahoo.com\\Correspondence to : S. Chatterjee} \\[6mm]

\end{center}
\begin{abstract}
Following the wellknown spacetime decomposition technique as
applied to (d+1) dimensions we write down the equations of
magnetohydrodynamics (MHD) in a spatially flat generalised FRW
universe. Assuming an equation of state for the background cosmic
fluid we find solutions in turn for acoustic waves and also for
Alfven waves in a warm (cold) magnetised plasma. Interestingly the
different plasma modes closely resemble the flat space
counterparts except that here the field variables all redshift
with their time due to the expansion of the background. It is
observed that in the ultrarelativistic limit the field parameters
all scale as the free photon. The situation changes in the
prerelativistic limit where the frequencies change in a bizarre
fashion depending on initial conditions. It is observed that for a
fixed magnetic field in a particular medium the Alfven wave
velocity decreases with the number of dimensions, being the
maximum in the usual 4D. Further for a fixed dimension the
velocity attenuation is more significant in dust compared to the
radiation era. We also find that in an expanding background the
Alfven wave propagation is possible only in the high frequency
range, determined by the strength of the external magnetic field,
 the mass density of the medium and also the dimensions of the
spacetime. Further it is found that with expansion the cosmic
magnetic field decays more sharply in higher dimensional
cosmology, which is in line with observational demand.
\end{abstract}
   ~~~~~~~KEYWORDS : cosmology; higher dimensions; plasma

~~~PACS :   04.20, 04.50 +h

\section*{1. Introduction}

The present investigation is a continuation of our earlier work
~\cite{jcap} where we studied the propagation of an
electromagnetic wave in a magnetised(un) plasma medium in an
expanding universe. So we will be rather brief in mathematical
details as also in other presentations. The work is primarily
motivated by two factors. Although plasma is the most significant
component of the cosmic fluid, barring some disjointed works the
much sought after union between the plasma dynamics and cosmology
has not been pursued so far to its logical end in a systematic
way. Secondly during the early phase of the universe before
recombination the cosmic fluid was presumably mostly in the form
of a plasma whereas in the cosmological context the higher
dimensional spacetime is particularly relevant in the early
universe. When considering the cosmic evolution it is believed
that during expansion at temperature  ($T > m_{e}$) the matter
existed mostly in the form of an electron positron plasma at ultra
relativistic temperature during $10^{-3}s<t<1s$. This is followed
by element formation so then we get essentially plasma of ionised
hydrogen in thermal equilibrium with radiation. At this stage the
energy density of radiation exceeded that of matter and the period
loosely termed radiation dominated era. With further cooling ($T <
m_{e}$) due to expansion recombination of ionised plasma took
place so that matter got decoupled from radiation and we call it
matter dominated era. On the other hand it can be shown that
Einstein's equations generalised to higher dimensions ~\cite{hoop}
do admit solutions where as time evolves the `extra' dimensions
shrink in size to be finally invisible with the current state of
experimental techniques so the present visible world is
effectively four dimensional. So we have thought it fit to study
some of the plasma processes in an expanding background in the
framework of higher dimensional spacetime. And this we have
attempted to do in a systematic way through the assumption of an
equation of state (EOS) for the background fluid such that both
the ultrarelativistic and prerelativistic  era can be dealt with
in a rather unified manner. We here tacitly assume that the plasma
field is weak enough not to back react on the cosmic expansion,
which is determined by the EOS obtaining at the particular
instant. Moreover both the plasma field equations and the GR being
highly nonlinear we are forced to consider a linearised plasma
model for mathematical convenience.\vspace{0.5cm}

As commented earlier scant attention has been paid so far for the
formulation of equations of  magnetohydrodynamics (MHD) in curved
spacetime, mathematical complexity being a strong deterrent in all
the approaches. To our knowledge, Holcomb and
Tajima~\cite{HT,holc}, Dettmann etal~\cite{det}, Banerjee
etal~\cite{sil,ab} are some of the authors making significant
contributions in this regard. Like our previous work here also we
adopt the (d+1) decomposition of spacetime as originally
formulated by MacDonald and Throne ~\cite{Mac}. In this approach
there exists a set of preferred fiducial observers with respect to
which the field quantities like mass density, velocity and
electromagnetic field are all decomposed and measured. To
illustrate the effects of curvature and expansion on MHD results
we consider the spatially flat FRW metric generalised to (d+1)
dimensions as the simplest background available. At some stage we
also assume the existence of an external magnetic field small
enough not to disturb the isotropy of the FRW background i.e., we
consider the region of space smaller than the horizon where the
magnetic field may be taken to be coherent. In fact magnetic
fields have been detected and measured in our galaxy and in our
local group with the help of Zeeman splitting as also Faraday
rotation measurements of linearly polarised radio waves from
pulsar sources. We find that with expansion the cosmic magnetic
field dissipates more efficiently in HD cosmology compared to the
4D one. Given the fact that our cosmology is manifestly
homogeneous and isotropic this rapid decay is desirable from the
observational point of view. In a generalised FRW type of metric
we systematically discussed  the
 propagation of photons in empty space, in both cold and warm plasma
 with or without external magnetic field and find exact solutions of
 analogous equations of motion. Though not specifically
 stressed in the previous works by others we find that the different
 asymptotic modes closely resemble their flat space counterparts except
 by a conformal factor, which unlike the Newtonnian case makes the frequencies
 time dependent. Interestingly all the modes scale exactly as the free photon
 case except the matter dominated magnetised modes where the Alfv\'{e}n
 wave frequencies redshift in a bizarre fashion, depending on the mass density
 of the medium, strength of the magnetic field and the number of dimensions.
 We also find that the Alfv\'{e}n waves can either be evanescent or propagating
 depending on some initial conditions. It is found that the medium allows
 only high frequency waves to pass and with increase in dimensions the cutoff
frequency also increases.\vspace{0.5cm}

We proceed to tackle our problem as follows: section 2 deals with
the (d+1) decomposed GRMHD equations in the FRW setting while in
section 3 we discuss the accoustic and Alfv\'{e}n waves and its
characteristics. The work ends with a discussion in section 4.

\section*{2.  MHD equation in an expanding universe}

 We extend here the (3+1) decomposition of GR as
    formulated by Arnowitt, Deser and Misner (ADM) ~\cite{adm} to a
    higher dimensional space time of (d+1) dimensions.  As the split
    formalism has been extensively discussed and used in the literature
~\cite{HT,holc} we skip the formalities and proceed to write down
the final GRMHD equations for a spatially flat generalized FRW
metric

\begin{equation}
ds^{2} = dt^{2} - A^{2} \left( dx^{2} + dy^{2} + dz^{2} +
d\psi_{n}^{2}\right)
\end{equation}
(n = 5, 6, 7,  \ldots , d )

where $A \equiv A (t)$ is the scale function. In a recent
communication\cite{jcap} we discussed in some detail the plasma
phenomena in generalised FRW universe in the absence of an
external magnetic field. We would like to complete our search by
considering a magnetised plasma in this section to eventually find
an expression for the Alfven wave. Let an ambient, large scale
magnetic field ~\cite{chen} be present in the plasma medium and
assuming a perfect fluid approximation we study the MHD equations
in a curved background.\vspace{0.5cm}

We assume that the homogeneous magnetic field `B' is generated
prior to the time of recombination. Such a field could generated
during electroweak phase transition ~\cite{giov}.We write
\begin{equation}
B = B_{0}(t) + B_{1}(x,t)
 \end{equation}

where $B_{1}$ denotes the first order perturbation to the magnetic
field. In an earlier work one of us ~\cite{mnras} explicitly found
the solutions of the equations of motions for this spacetime via
an equation of state, $P = \alpha \rho$ ( $P$ = pressure, $\rho$ =
energy density) such that the scale factor is
\begin{equation}
A \sim t^{\frac{2}{d(1+\alpha)}} = t^{n},~~~ n =
\frac{2}{d(1+\alpha)}
\end{equation}
With the extrinsic curvature scalar defined as
\begin{equation}
K = - \theta = - d\frac{\dot{A}}{A} = - d\frac{n}{t}
\end{equation}

we finally write down the curl equations of Maxwell ~\cite{Mac}
(see Mcdonald et al for ( 3+1 ) split for more details )
generalized to ( d+1) dimensions as

\begin{eqnarray}
  A^{(d-1)}(\nabla \times E) = - \frac{1}{c}\frac{\partial}{\partial
t}(A^{d}B)
\end{eqnarray}
and
\begin{eqnarray}
  A^{(d-1)}(\nabla \times B) =  \frac{1}{c}\frac{\partial}{\partial
t}(A^{d}E)+ \frac{4\pi j}{c}A^{d}
\end{eqnarray}
where the symbols have the usual significance. Again assuming a
perfect conducting fluid for which the resistive force will be
zero, we write
\begin{eqnarray}
 E = - A \frac{v}{c}\times B
\end{eqnarray}

Now the equation of charge conservation is given by,
\begin{eqnarray}
 \frac{d}{dt}(A^{d}\rho _{e}) = - A^{d}\nabla.j
\end{eqnarray}
    We assume that a perfect gas adiabatic EOS holds for a (d+1)
    dimensional case also such that $P = \sigma
\rho^{\gamma}$ where   $\gamma$ is an adiabatic constant.
Following Wienberg  ~\cite{wien} one can easily show that here
$\gamma$ comes out to be  $\gamma = \frac{d+1}{d}$ and the
relativistic enthalpy,
\begin{eqnarray}
h = 1 + \frac{U}{\rho c^{2}} + \frac{P}{\rho c^{2}}
\end{eqnarray}
Further pressure gradient , $\nabla P = hC_{s}^{2}\nabla\rho$
where the velocity of sound is given by $C_{s}^{2}= \frac{\gamma
P}{\rho h}$. Now the linearised HD force balance equation is given
by,

\begin{eqnarray}
\frac{\partial}{\partial t}\left[t^{n(2d-1)}\rho_{0}hv_{1}\right]
= t^{2nd}\frac{j\times B}{c} - t^{nd}h C_{s}^{2}\nabla \rho_{1}
\end{eqnarray}
 For both radiation and matter dominated cases $\rho_{0}h \sim t^{-2}$
 and $h C_{s}^{2} \sim t^{-nd(\gamma -1)}$ ($\rho_{0}$ = mass density). From
 equation (6), neglecting displacement current, we get,
\begin{eqnarray}
t^{n}j = \frac{c}{4\pi}\left(\nabla \times B \right)
\end{eqnarray}
Using the above relations we get from HD force balance equation,
\begin{eqnarray}
&&\frac{\partial}{\partial t}\left[t^{nd(\gamma -
1)}\frac{\partial}{\partial t}
 \left\{t^{n(2d-1)-2}v_{1}\right \} \right]\nonumber
 \\
 =&-&\frac{1}{4\pi} \left[n \{d(\gamma - 1) -1 \}\right]t^{n\{d(\gamma
- 1)-1 \}-1}t^{nd}B\times (\nabla \times
t^{nd}B) \nonumber\\
&-& \frac{1}{4\pi} t^{n\{d(\gamma - 1) -1
\}}\frac{\partial}{\partial t}\left(t^{nd}B \right)\times \nabla
\times \left(t^{nd}B \right)\nonumber\\
 &-& \frac{1}{4\pi}
t^{n\{d(\gamma - 1) -1 \}}t^{nd}B \times \nabla \times
\frac{\partial}{\partial t}\left(t^{nd}B \right)-
\frac{\partial}{\partial t} \left(t^{nd}\nabla \rho_{1} \right)
\end{eqnarray}

Using equation (2) we get,  $ \frac{\partial}{\partial t}(t^{nd} B
) \times \nabla \times t^{nd} B = 0 $ \vspace{0.5cm}
 But,
\begin{eqnarray}
t^{nd}B \times \nabla \times \frac{\partial}{\partial t} (
t^{nd}B) = t^{2nd} B_{0} \times \nabla \times \nabla \times v_{1}
\times B_{0}
\end{eqnarray}
and
\begin{eqnarray}
\nabla \frac{\partial}{\partial t}(t^{nd} \rho_{1}) = -
t^{nd\gamma} hC_{s}^{2} \rho_{0} \nabla (\nabla.v_{1})
\end{eqnarray}
Using the above equation we get,
\begin{eqnarray}
&&t^{2}\frac{\partial^{2}v_{1}}{\partial t^{2}} + \left(nd\gamma +
3nd -2n -4 \right) t \frac{\partial v_{1}}{\partial t}
\nonumber\\&+& (2nd - n
- 2)( nd\gamma + nd + -n - 3)v_{1} \nonumber\\
 &+& t^{4 -
2nd}v_{Ai}\times \nabla \times \nabla \times v_{1} \times
v_{Ai}\nonumber\\& -& t^{-(nd\gamma + nd -n - 4)}C_{si}^{2} \nabla
(\nabla.v_{1}) = M_{i}n(d\gamma - d - 1) t^{n+3-2nd}v_{1} \times
v_{Ai}
\end{eqnarray}

where, $v_{Ai} = \frac{t^{nd-1}B_{0}}{(4\pi \rho_{0}
h)^{\frac{1}{2}}}$, is a time invariant vector and the suffix
$`i$' denotes an arbitrary initial instant.

\hspace{1.2cm} $C_{si}^{2} = t^{nd\gamma-2} C_{s}^{2} =
t^{\frac{2(1-d\alpha)}{d(1+\alpha)}}C_{s}^{2}$

\hspace{1.2cm} $M_{i} = \sqrt{\frac{4\pi \rho_{e0}^{2}}{\rho_{0} h
c^{2}}}t^{nd - 1}$

It has not escaped our notice that the right hand side (RHS) of
the equation (15) disappears whenever we consider an
ultrarelativistic plasma in higher dimensional space time i.e.,
$\gamma = \frac{d+1}{d}$, irrespective of the value of n, the
expansion rate of the back ground FRW metric. The term
$M_{i}n(d\gamma - d - 1)t^{n+3-2nd}v_{1}\times v_{Ai}$ remained
unnoticed in the work of Holcomb et al.~\cite{HT} but it was first
noticed in the work of Sil et al ~\cite{sil}. This term will
disappear for ultrarelativistic case only.\vspace{0.5cm}

As in Newtonian mechanics the relativistic Alfv\'{e}n wave
velocity $v_{A}$ is defined as
\begin{eqnarray}
v_{A} = \frac{B_{0}}{\sqrt{4\pi \rho_{0} h}} = v_{Ai}t^{1-nd} =
v_{Ai}t^{-\frac{1-\alpha}{1+\alpha}} \end{eqnarray}

 where
$v_{Ai}$ is the invariant Alfv\'{e}n wave velocity at any
instant~$t = t_{i}$. For the prerecombination era in
(d+1)-dimensional spacetime, $\alpha = \frac{1}{d}$ for perfect
fluid and then $v_{A} \sim t^{-\frac{d-1}{d+1}}$. Hence the
Alfv\'{e}n velocity depends upon dimensions and decreases with it.
It is maximum at 4D. So the Alfv\'{e}n velocity actually decreases
in higher dimensions. \vspace{0.5cm}

On the other hand for a fixed $`d$' the Alfv\'{e}n velocity varies
as~ $t^{-1}$ for the postrecombination era ($\alpha = 0$).
Alternatively, for the case of very large number of dimensions,
the damping asymptotically reaches $t^{-1}$, a form set for the
postrecombination era also. It is to be noticed that the velocity
for $\alpha = 0$ is independent of the number of dimensions,
unlike the radiation case. The Alfv\'{e}n wave velocity decreases
more sharply with time in dust (see figure 1). \vspace{0.5cm}


\noindent
 From equation (5) and (7) it follows that
$\frac{\partial}{\partial t}(A^{d}B_{0}) = 0$ ; or,
$\frac{\partial}{\partial t}(t^{\frac{2}{1+\alpha}}B_{0}) = 0$.
For radiation case $(\alpha = \frac{1}{d})$, $(B_{0})_{rad} \sim
t^{-\frac{2d}{d+1}}$ with respect to comoving co-ordinates.
\vspace{2cm}
\begin{figure}[ht]
\begin{center}
 \includegraphics[width=8cm]{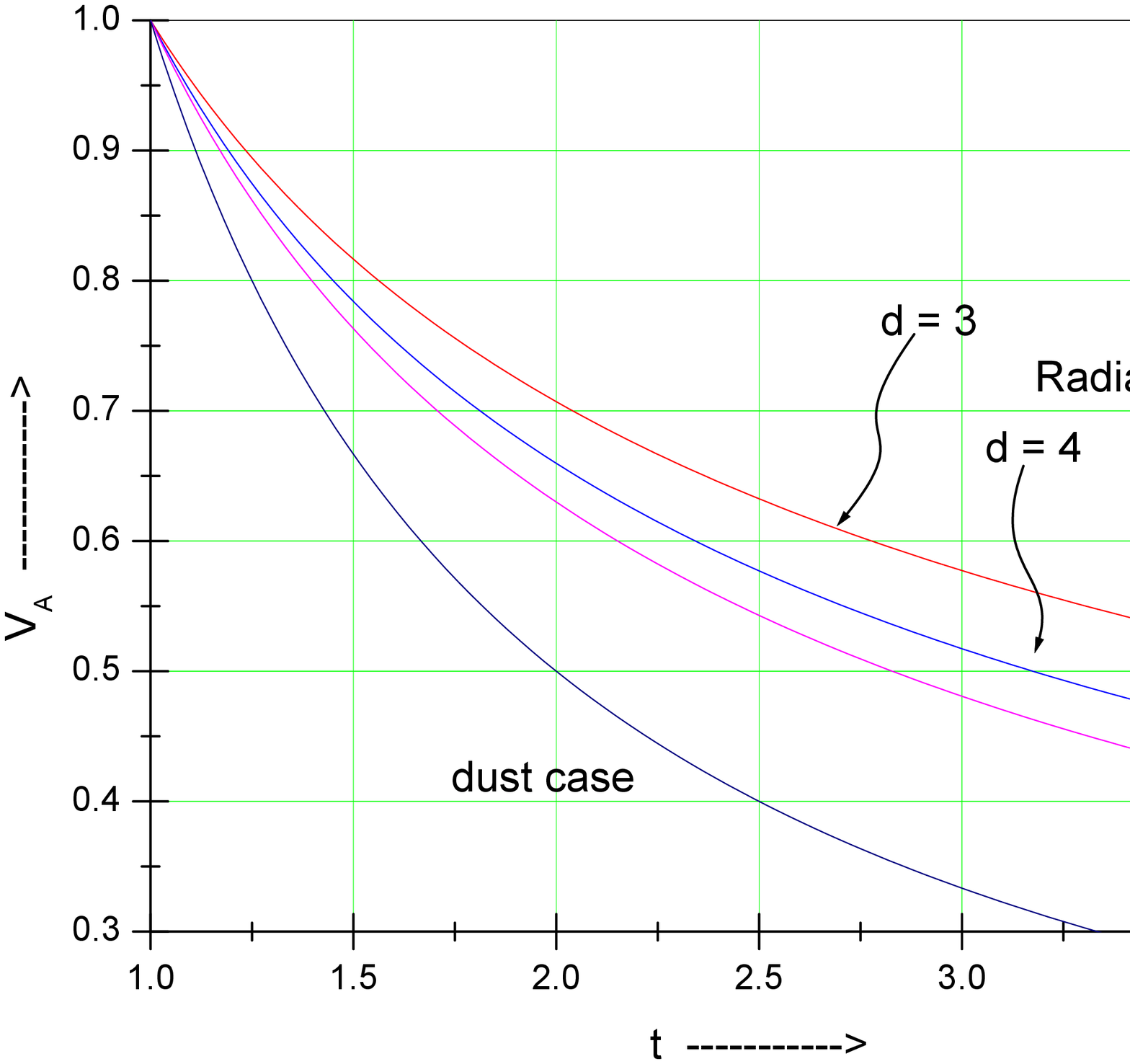}
  \vspace{-2.5 cm}
  \caption{
 \small \emph{ $v_{A} \sim t  $ graph}
}
\end{center}
\end{figure}

 It was
pointed before that $\rho_{0}h \sim t^{-2}$  both in higher
dimensions and 4D and also in radiation and dust case.
 For the prerecombination era in (d+1)- dimensional
spacetime $B_{0}$ decreases with increasing dimensions, being
 maximum in 4D.  Again for the postrecombination era (i.e,
$\alpha =0$) the ambient magnetic field $B_{0} \sim t^{-2}$ for a
fixed $`d$'. $B_{0}$ reaches asymptotically $t^{-2}$ for a very
large number of dimensions in radiation case and the nature is
same as the postrecombination era (see figure 2).

\vspace{0.3 cm}
\begin{figure}[h]
\begin{center}
\vspace{2.2 cm}
 \includegraphics[width=8cm]{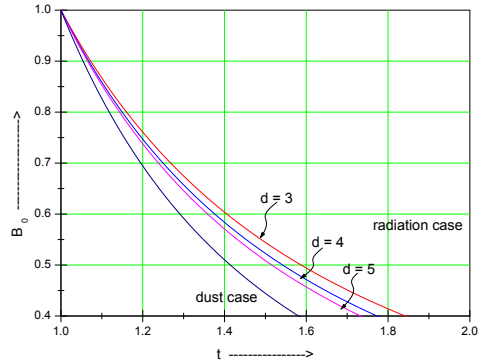}
\vspace{-2.5 cm} \caption{
\small \emph{ $B_{0} \sim t  $ graph}
}
\end{center}
\end{figure}
 \noindent
The fact that $v_{A}$ decreases in an expanding background owes an
explanation as follows : we know from classical idea that
Alfv\'{e}n wave is generated as in a transverse vibration in a
string as an interplay between the unbalanced tension and inertia.
When looked upon as lines of force the tension in the string comes
out to be $\frac{B^{2}}{8\pi}$, the energy density of the magnetic
field. Since with higher dimensions the volume increases the
energy density, and consequently the magnetic field B also gets
diluted. So the Alfv\'{e}n velocity also decreases with increasing
dimensions. \vspace{0.5cm}

One can look at from a different standpoint also. As a consequence
of flux conservation we may write

\begin{eqnarray}
\dot{\Phi} = \frac{d}{dt}\int A^{d}B ds = \frac{d}{dt}\int
t^{\frac{2}{1+\alpha}} = - \varepsilon ~~~(e.m.f.)
\end{eqnarray}

Since the uniform ambient magnetic field does not produce an e.m.f
we see that if $\varepsilon = 0$, as  the magnetic field must
decay in time $t^{-\frac{2}{1+\alpha}}$ with respect to co-moving
co-ordinates. This result is also same for 4D case as well as for
higher dimensions. Thus as a consequence of magnetic flux
conservation the field lines in an expanding universe are
conformally diluted.\vspace{0.5cm}

The theoretical result that with expansion the cosmic magnetic
field decays much faster in HD compared to 4D may have non trivial
implications in astrophysics. In a recent communication Gang Chen
etal.~\cite{pia} argued that if a primordial magnetic field did
exist it should create vorticity and alfv\'{e}n waves and finally
cosmic microwave background (CMB) anisotropies. They later went to
discuss the constrints that can be placed on the strength of such
a field with the help of the CMB anisotropy data from WMAP
experiment. The origin of the observed magnetic galactic field of
a few $\mu G$ apparently coherent over a 10 kpc scale continues to
evade satisfactory theoretical explanation. A plausible
explanation may be the consequence of nonlinear amplification of a
tiny seed field by galactic dynamo process~\cite{zwei}. CMB
anisotropy data apparently suggests a seed field of $nG$
~\cite{pia} strength. Now the dynamo mechanism being so efficient
vis-a-vis amplification of the seed field at the early era it is
quite likely to shoot up the seed field beyond the current value
of a few $\mu
 G$. However that may introduce considerable anisotropy~\cite{zeld}
  in the manifestly homogeneous and isotropic FRW model. And here the
 HD model makes its presence felt in allowing the amplified field to decay much
faster.
\vspace{0.5cm}

Now, putting $\gamma = \frac{d+1}{d}$ and $n =
\frac{2}{d(1+\alpha)}$ in equation (15) we get finally,
\begin{eqnarray}
t^{2}\frac{d^{2}v_{1}}{dt^{2}} + \frac{4d-4d\alpha-2}{d(1 + \alpha
)}t\frac{dv_{1}}{dt} + \nonumber
\frac{(2d - 2d\alpha - 2 )(1 - 3\alpha)}{d(1 + \alpha)^{2}}v_{1}~~~~~ \\
+ ~ t^{\frac{4\alpha}{1+\alpha}}v_{Ai}\times \nabla \times \nabla
\times v_{1} \times v_{Ai} -
t^{\frac{4\alpha}{1+\alpha}}C_{si}^{2}\nabla(\nabla.v_{1}) = 0
\end{eqnarray}

 The above equation is quite complex. It can be simplified,
however, if we consider the interesting case of the wave moving
perpendicularly to both magnetic field and wave vector. In this
case

\begin{eqnarray}
v_{Ai}\times \nabla \times (\nabla \times v_{1} \times v_{Ai}) =
k_{i}^{2}v_{Ai}^{2}v_{1}
\end{eqnarray}
and

\begin{eqnarray}
C_{si}^{2}\nabla(\nabla.v_{1}) = - C_{si}^{2}k_{i}^{2} v_{1}
\end{eqnarray}

so the above equation gives
\begin{eqnarray}
t^{2}\frac{d^{2}v_{1}}{dt^{2}} + \frac{4d-4d\alpha-2}{d(1 + \alpha
)}t\frac{dv_{1}}{dt}\nonumber
~~~~~~~~~~~~~~~~~~~~~~~~~~~~~~~~~~~~~~~~~~~~~~~~
 \\  + \left[\frac{(2d - 2d\alpha - 2 )(1 -
3\alpha)}{d(1 + \alpha)^{2}} +
t^{\frac{4\alpha}{1+\alpha}}k_{i}^{2}(v_{Ai}^{2}+
C_{si}^{2})\right]v_{1} = 0
\end{eqnarray}
 where the velocities,$C_{si}$ originates from the acoustic
 pressure and $v_{Ai}$ from the magnetic field. This is a Bessel
 type of equation of order, $p = \frac{d + d\alpha -
 2}{4d\alpha}$~ $(\alpha \neq 0 )$.

The equation (21) yields a general solution

\begin{eqnarray}
v_{1} = t^{\frac{-3d + 5d\alpha + 2}{2d
(1+\alpha)}}H_{(p)}\left[(v_{Ai}^{2} +
C_{si}^{2})k_{i}\frac{1+\alpha}{2\alpha}t^{\frac{2\alpha}{1+\alpha}}\right]
\end{eqnarray}

 \vspace{0.5cm}
 \textbf{Special Cases:} We now briefly discuss the
equation (22) for the special cases of ultra relativistic
($T>>m_{c}$) and also stiff fluid back ground. Evidently the
prerelativistic case, $\alpha = 0$ is not obtainable from the
above equation and needs a separate treatment. So we take
radiation and for the sake of completeness the stiff fluid case
also.\\

 \textbf{ I. Radiation Case :} Putting $\alpha =
\frac{1}{d}$  the equation (22) reduces to
\begin{eqnarray}
v_{1} =
t^{-\frac{(3d-7)}{2(d+1)}}H_{(p_{1})}\left[\frac{d+1}{2}(v_{Ai}^{2}
+ C_{si}^{2})k_{i}t^{\frac{2}{d+1}} \right]
\end{eqnarray}
 where the order of the Henkel coefficient is given by
 $p_{1} = \frac{d-1}{4}$, the frequency of the oscillation
 $\sim t^{-\frac{d-1}{d+1}}$ and $(v_{Ai}^{2} +
 C_{s}^{2})^{\frac{1}{2}}$ is known as phase velocity. This mode
 corresponds to the magnetosonic or fast Alfv\'{e}n mode because
 both velocities are simultaneously present. The result is identical
 to the flat space case also except that frequency redshifts here and is model
 dependent. \vspace{0.5cm}

 This wave is somewhat similar to an electromagnetic wave, since
 the time varying magnetic field is perpendicular to the direction
 of propagation but parallel to the magnetostatic field, whereas
 the time varying electric field is perpendicular to both the
 direction of propagation and the magnetostatic  field which is a
 longitudinal wave. However, since the velocity of mass flow and
 also fluctuating mass density associated with the wave motion are both
 in the wave propagation direction this wave is called the magnetosonic
  longitudinal wave. The phase velocity $(v_{Ai}^{2} +
 C_{s}^{2})^{\frac{1}{2}}$ is independent of frequency so that it is a non
 dispersive wave. We see that the fluid velocity depends upon
 dimension also. It will reduce to Holcomb et al ~\cite{HT} work for $d =
 3$. As noted earlier the attenuation of the wave propagation is
 more prominent in the higher dimensional regime compared to the
 4D case$ \left(v_{1} \sim t^{\frac{-3d+5}{2(d+1)}}\right)$. As
 the dimension `$d$' increases indefinitely the
 damping finally becomes independent of `$d$', $d \rightarrow
 \infty$ and $v_{1} \sim t^{-\frac{3}{2}}$~(see figure 3).
\begin{figure}[ht]
\begin{center}
\vspace{2.2 cm}
 \includegraphics[width=8cm]{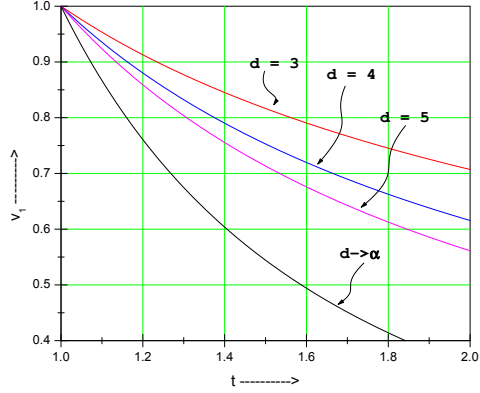}
\vspace{-2.5 cm}
\caption{
\small \emph{ $v_{1} \sim t  $ graph}
}
\end{center}
\end{figure}
 Now if one can consider a relatively cold plasma such that the
  `pressure' due to magnetic field far exceeds the acoustic pressure one may take
   $C_{si} = 0$, this mode is known as pure or shear
 Alfv\'{e}n wave. In this case only magnetic field is present, but the
 nature of wave will be same as magnetosonic wave.   Another
 interesting feature comes out of the analysis. We have shown
 earlier ~\cite{jcap} like others in the related field that a free
 photon redshifts  in the ultrarelativistic case and the field
 variables in the different plasma models also redshift
 identically. We get exactly similar result for the Alfv\'{e}n
 mode also. Though not specifically emphasized in the earlier
 works this result, in our opinion, is not model independent. The
 fact that a conformally flat FRW mechanism is chosen in al our
 works all the field variables here scale with the conformal scale
 factor. A less symmetrical spacetime would possibly yield different
 results. We shall, however, subsequently see that the situation
 drastically changes in the non relativistic background.

 On the other hand for very low magnetic field the acoustic
 pressure dominates and we get pure acoustic wave.
 \vspace{0.5cm}

\textbf{ II. Stiff Fluid} ($\alpha = 1$)\textbf{:} For the sake of
completeness we also consider the
 stiff fluid case  characterised by $\alpha = 1$. Equation
 (22) gives ($C_{si}$= c = velocity of light ).
\begin{eqnarray}
v_{1} =
t^{\frac{d+1}{2d}}H_{\left(\frac{d-1}{2d}\right)}\left[(v_{Ai}^{2}
+ c^{2})k_{i}t \right]
\end{eqnarray}

However, the stiff fluid being not  much physically relevant we
skip discussion of this result.\vspace{0.5cm}

\textbf{III. Dust Case} ($\alpha = 0$)\textbf{:} Here equation
(21) becomes,

\begin{eqnarray}
t^{2}\frac{d^{2}v_{1}}{dt^{2}} + \frac{4d-2}{d} t\frac{dv_{1}}{dt}
+ \left[\frac{2(d-1)}{d} + k_{i}^{2}v_{Ai}^{2}\right]v_{1} = 0
\end{eqnarray}

It is clear that the form of the two equations corresponding to
radiation and dust cases are distinctly different - one resembling
the Bessel form while the other to an Euler type. In this case if
the magnetic field is very strong so that the magnetic pressure is
very much larger than the fluid pressure, then phase velocity
 of the magnetosonic wave becomes equal to the
Alfv\'{e}n wave velocity $v_{Ai}$. The second one has three
solutions depending upon the discriminant $D = (d-2)^{2} - \left(2dk_{i}^{2}v_{Ai}^{2}\right)$:

\vspace{0.5cm}
 (a)~For $D > 0$,
\begin{eqnarray}
v_{1} = c_{1}t^{-\frac{1}{2d}[(3d-2)+\sqrt{D} ]} +
c_{2}t^{-\frac{1}{2d}[(3d-2)-\sqrt{D} ]}
\end{eqnarray}
where $c_{1}$ and $c_{2}$ are constants. This is a general
solution of equation (21). Here the above mode always dies out
without much propagation.\vspace{0.5cm}

(b)~For $D = 0$,
\begin{eqnarray}
v_{1} = \left(c_{3} + c_{4}\ln t \right)t^{-\frac{(3d-2)}{2d}}
\end{eqnarray}
$c_{3}$ and $c_{4}$ are constants. It is apparent that $v_{1}$is
least in the usual 3D. This result is somewhat intriguing when
posited against the fact that the phenomena of damping is due to
the background expansion, which is maximum in 3D. So other process
may be in work besides the expansion rate in this
case.\vspace{0.5cm}

 (c)~ For $D< 0$,
\begin{eqnarray}
v_{1} = c_{5}t^{-\frac{3d-2}{2d}} \cos
\left(\frac{\sqrt{D}}{2d}\ln t \right) + c_{6}t^{-\frac{3d-2}{2d}}
\sin \left(\frac{\sqrt{D}}{2d}\ln t \right)
\end{eqnarray}
where $c_{5}$ and $c_{6}$ are constants. So we get damped
oscillatory mode here, although the periodicity is not in cosmic
time but in its logarithm. Hence the signature of of D which, in
turn, depends on the dimension, external magnetic field and the
density of the medium is crucial in determining whether a
particular mode will be evanescent or propagating. In fact for
oscillation to take place one should have $k_{i}v_{Ai} >
\frac{d-1}{d}$. Now for dust $v_{Ai}$ is independent of $`d$'. So
as number of dimension increase only very high frequency waves can
propagate through a particular plasma medium. The extra
dimensions, so to say, try to filter the frequency range. This
finding is interesting but its astrophysical implications are yet
to be looked into. The shear Alfv\'{e}n waves are stationary in a
certain frequency range which depends upon D. For higher
frequencies, the waves do propagate; however, they are not
oscillatory in the cosmic time, but in its logarithm.

\section*{3. Discussion}
\vspace{0.5cm}
 As physics in higher dimension is increasingly becoming an
 important and at the same time a distinct area of activity we are primarily
 interested in this work to investigate the effect of higher dimensions in the
  propagation of an electromagnetic
  wave in plasma media with an expanding background. As commented earlier
  in the introduction this work may be looked upon as continuation of our
  earlier work where acoustic and Alfv\'{e}n modes are not considered. Here the
  background cosmology chosen is the simplest possible - FRW
  spacetime generalized to (d+1) dimensions. So the results we
  obtain differ essentially in some quantitative aspects without
  introducing much qualitative differences.\vspace{0.5cm}

   The interesting results may be briefly summarised as follows: We know that the
  expanding background generates a sort of damping such that the
  plasma variables like the amplitude of the wave, strength of the magnetic
  field etc. decay. Unlike mechanical damping the expansion
  creates a thinning of density of magnetic lines of force so that the
  velocity of the different modes keep on decreasing with time. Naturally for a
  contracting mode just the opposite will happen. We observe that this
  damping is the least in the usual 3D space compared to the HD space.
  As commented in the introduction this exercise may be looked upon
 as continuation of our earlier work where acoustic or Alfv\'{e}n
 waves are not discussed. The locally measured frequencies resemble
 that of flat spacetime, though due to the expansion of the
 universe they do decay in time. We have found like others in the related
 field that in the ultra relativistic limit all modes redshift in
 the same scale as free photon. In the non relativistic limit,
 however, the situation drastically changes. Different modes redshift at
 different rates depending on the plasma density, magnitude of the
 external magnetic field and interestingly on number of dimensions. So
 here dimensions may embed a definite signature. These factors also
 dictate whether a particular mode becomes evanescent or can propagate  through the medium.
 We find that as dimension increases only very high frequency
 waves can propagate but they are not oscillatory in cosmic time but in its logarithm.
Finally coming back to the vexed question of a primordial magnetic
field we find the field decays much faster in HD spacetime, which
is a good news for the observed isotropy and homogeneity of the
present universe because large magnetic field introduces
anisotropy.\vspace{0.5cm}

To end a final comment may be in order. We know that at the early
phase when consideration of both higher dimensions
  and plasma phenomena are interesting the fluid was far from being perfect.
  Viscosity and heat processes were significant.  Possible extension to this work
 could include formulating these equations for imperfect MHD
 including such effects as viscosity and finite conductivity.
 Nonlinear modes such as shocks could also be investigated.
 Judging by the complexity of the solutions it seems likely that
 more complicated modes may not be analytically solvable and may
 require extensive numerical analysis. Our simple analysis has
 kept intact the basic essence of MHD theory, highlighting the
 similarities and occasional differences with standard flatspace
 MHD, which complicated numerical analysis may mask.

\vspace{0.5cm}

\textbf{Acknowledgment : } One of us(SC) acknowledges the
financial support of UGC, New Delhi for a MRP award.

\end{document}